\documentclass[twocolumn,prb,showpacs,superscriptaddress]{revtex4}
\usepackage{epsfig}
\begin{document}
\title {
 Electronic Structure and Linear Optical Properties of Sr$_{2}$CuO$_{2}$Cl$_{2}$ Studied from the First Principles Calculation
}
\author{Hongming Weng}\email[Corresponding author E-mail:]{silicon@nju.edu.cn}
\affiliation {Group of Computational Condensed Matter Physics,
National Laboratory of Solid State Microstructures and Dept. of
Physics, Nanjing University, Nanjing 210093, P.R.China}
\author{Xiangang Wan, Jian Zhou and Jinming Dong}
\affiliation {Group of Computational Condensed Matter Physics,
National Laboratory of Solid State Microstructures and Dept. of
Physics, Nanjing University, Nanjing 210093, P.R.China}
\date{\today}
\begin{abstract}
First-principles calculations with the full-potential linearized
augmented plane-wave (FP-LAPW) method have been performed to
investigate detailed electronic and linear optical properties of
Sr$_{2}$CuO$_{2}$Cl$_{2}$, which is a classical low-dimensional
antiferromagnet (AFM) charge transfer ({\it CT }) insulator.
Within the local-spin-density approximation (LSDA) plus the
on-site Coulomb interaction $U$ (LADA+$U$) added on Cu $3d$
orbitals, our calculated band gap and spin moments are well
consistent with the experimental and other theoretical values. The
energy dispersion relation agrees well with the angle resolved
photoemission measurements. Its linear optical properties are
calculated within the electric-dipole approximation. The
absorption spectrum is found to agree well with the experimental
result.
\end{abstract}

\vskip 2cm

\pacs{71.15.Ap, 74.25.Jb, 74.25.Gz}
 \maketitle
\section{introduction} \label{introduction}
The study of the strongly correlated electronic systems in two
dimensions (2D) has been a very active and interesting problem
since the discovery of the high $T_{c}$ superconductors, in which
calculation of the electronic structures of these materials by the
first principles methods is a first step toward a microscopic
understanding of them. Among these materials, the
Sr$_{2}$CuO$_{2}$Cl$_{2}$ is an idea paradigm since it has an
extremely weak coupling between its CuO$_{2}$ planes, and is thus
a more exaggerated quasi-2D system than the most other
high-$T_{c}$ materials.

Experimental and theoretical efforts have been made to investigate electronic structures of the Sr$_{2}$CuO$_{2}$Cl$_{2}$. For example, Wells {\it et al.}~\cite{1} made a measurement on the angle-resolved photoemission spectroscopy (ARPES) above its N\'{e}el temperature $256.5\pm 1.5$K and got its energy band dispersion relations, which are in consistent with the $t-J$ model calculations along some k-paths but not all. To achieve more precise results, Kim {\it et al.}~\cite{2} managed successfully to do the ARPES experiments on the Sr$_{2}$CuO$_{2}$Cl$_{2}$ at 150K, which is below its N\'{e}el temperature, and found that it would be better to include the second and third nearest neighbor hopping terms, i.e., to use the $t-t'-t''-J$ model~\cite{3,4,5,6,7}, for well describing the observed phenomenon. The first principles calculation is an important theoretical method to get information on the ground state of materials. Hua Wu {\it et al.}~\cite{8} used the linear combination of atomic orbital (LCAO) method to calculate the electronic structure of the Sr$_{2}$CuO$_{2}$Cl$_{2}$ in an AFM structure. Their results give out reasonable density of states (DOS) to some extent, but do not show the dispersion relationship. D. L. Novikov {\it et al.}~\cite{9} also studied the Sr$_{2}$CuO$_{2}$Cl$_{2}$ by FP-LMTO numerical calculation without taking into account its magnetic structure and the strong correlation effect of Cu $3d$ electrons.

So, in this paper we try to do a more detailed study on the AFM Sr$_{2}$CuO$_{2}$Cl$_{2}$ with the FP-LAPW~\cite{10} method within the local-spin-density approximation and plus the on-site Coulomb interaction on Cu $3d$ orbitals. Then, based on this and within the electric-dipole approximation, its linear optical responses are also calculated. A lot of works~\cite{11,12,13,14,15}, mostly are model calculations and experimental measurements, had been done to study its optical properties for many years in order to understand the microscopic mechanism of the high-$T_c$ superconductivity exhibited in the layered copper oxides. As to our best knowledge, there is still no first-principles calculation study on its optical properties. Our result gives out a broad peak at the charge transfer ({\it CT}) gap, which is the dominant optical absorption feature in these materials, and other peaks in higher energy range, which are also well comparable with the experimental data.

\section{METHODOLOGY} \label{METHODOLOGY}

FP-LAPW~\cite{10} method treats all electrons without the shape
approximations for the potential and charge density. The
Perdew-Wang's exchange-correlation energies are used in the local
density approximation (LDA)~\cite{16}. In our calculation we have
taken the crystal and magnetic structures of the
Sr$_{2}$CuO$_{2}$Cl$_{2}$ from Ref. [17]. As shown in Fig. 1, in
the antiferromagnetic phase, the unit cell is doubled while the
first Brillouin zone (BZ) shrinks to half. The radii of the atomic
spheres are taken as 2.0 a.u. for Cu, Cl, and Sr and 1.6 a.u. for
O, respectively. And the cutoff of the plane-wave basis is set to
$R_{mt}K_{max}$=7.0 (the $K_{max}$ is the plane-wave cut-off, and
the $R_{mt}$ is the smallest radius for all atomic sphere radii.).
Thus, the resulting number of the plane waves used in our
calculation is about 4950 plus 190 local orbitals. Self-consistent
calculations are performed with 200 k-points in the first BZ by
the tetrahedron method. For checking the numerical convergence, we
increase k-points to 500 and find that the difference between the
two converged total energies is about 0.02 mRy. In order to
include the strong correlation effect of the Cu $3d$ electrons, we
use the LSDA+$U$ approach~\cite{18} with $U$=6.0 eV and $J$=125
meV. And the case of $U$=7.5 eV is also calculated for comparison.
Here the $U, J $ values are taken from Ref. [8], [19] and [20].
Convergence is assumed when the difference between the input and
output charge densities is less than 0.0001 e/(a.u)$^{3}$ and the
difference of total energies in the last two iterations is less
than 0.0001 Ry.
\begin{figure}
\centering
\includegraphics[width=0.47\textwidth]{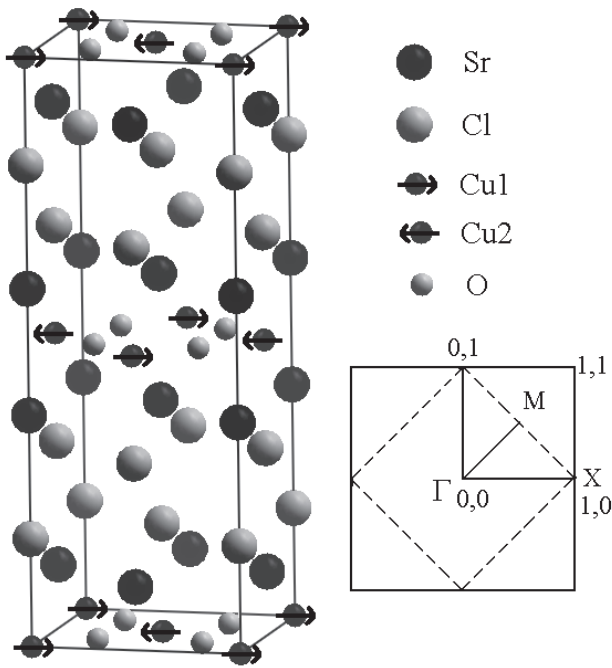}
\caption{The crystal and magnetic structure of the AFM
Sr$_{2}$CuO$_{2}$Cl$_{2}$ material. Cu1 means the Cu atom with
spin up and Cu2 spin down. Also shown is the two-dimensional
Brillouin zone. Solid (dashed) line is the one without (with)
consideration of the AFM magnetic structure.}\label{fig1}
\end{figure}
   Optical responses are calculated within the electric-dipole approximation. The imaginary part of the dielectric function can be expressed by
\[\begin{array}{l}
\varepsilon_2(\omega)=\frac{8\pi^2e^2}{\omega^2m^2V}
\sum\limits_{c,v} {\sum\limits_k {\vert < c,k\vert {\rm {\bf
\hat{e}}} \cdot } } {\rm {\bf p}}\vert v,k> \vert^2 \\
\times
\delta [E_c (k) - E_v (k) - \hbar \omega ], \\
\end{array}\]
\noindent
where $c$ and $v$ represent the conduction and valence bands, respectively, and $|c,k>$, $|v,k>$ are the eigenstates obtained from the FP-LAPW calculations. {\bf p} is the momentum operator, ${\rm {\bf \hat {e}}}$ is the external electric field vector, and $\omega$ is the frequency of incident photons. We still use 200 sampling k-points in the irreducible wedge and the linear tetrahedron scheme improved by Bl\"{o}chl {\it et al.}~\cite{21} for the BZ integrations. Using  Kramers-Kr\"{o}nig (K-K) transformation, we can get the real part of the dielectric function, and further calculate the absorption spectrum of the Sr$_{2}$CuO$_{2}$Cl$_{2}$.

\section{RESULTS AND DISCUSSION} \label{RESULTS AND DISCUSSION}

The calculated electronic structures for the AFM
Sr$_{2}$CuO$_{2}$Cl$_{2}$ with $U$=6.0 eV are plotted in Fig. 2, 3
and 4. From Fig. 2, it is seen that there is an indirect band gap
of about 1.2 eV with the top of the valence band at M($\pi $/2,
$\pi $/2) point and the bottom of conduction band at X($\pi $, 0)
point, which is smaller than the experimental value of 1.9
eV~\cite{8}, but much closer to it than the value of 0.84 eV when
$U$= 5 eV in Ref. [8]. If a larger $U$=7.5 eV is selected and
other parameters are unchanged, an indirect gap of 1.68 eV is
obtained, which is very close to the value of 1.63 eV obtained by
the LCAO method~\cite{8}. The energy dispersion relation agrees
very well with the results in Ref. [1] and [2]. But, along the
direction from X($\pi$, 0) to $\Gamma$(0, 0) it is better than
that obtained from the $t-J$ model calculation~\cite{1}. The
energy dispersion of the top-most valence band along $\Gamma$(0,
0) to M($\pi$/2, $\pi$/2) reaches to about 0.45 eV, larger than
the experimental value of 280$\pm $60 meV~\cite{1}. But
calculation with $U$=7.5 eV shows that a larger $U$ will suppress
the dispersion to 0.32 eV, which is much closer to the
experimental value. Anti-ferromagnetic order of
Sr$_{2}$CuO$_{2}$Cl$_{2}$ can be clearly seen from Fig. 3(a) and
3(b). The total spin is zero because both of Cu1 and Cu2 have
anti-parallel spin moments of about 0.59 $\mu _{B}$ per atom,
which is also bigger than the experimental value 0.34 $\mu
_{B}$~\cite{17}. It is found that the larger $U$ of 7.5 eV will
increase the spin moment of Cu atom to 0.63 $\mu _{B}$ per atom.

\begin{figure}
\centering
\includegraphics[width=0.47\textwidth]{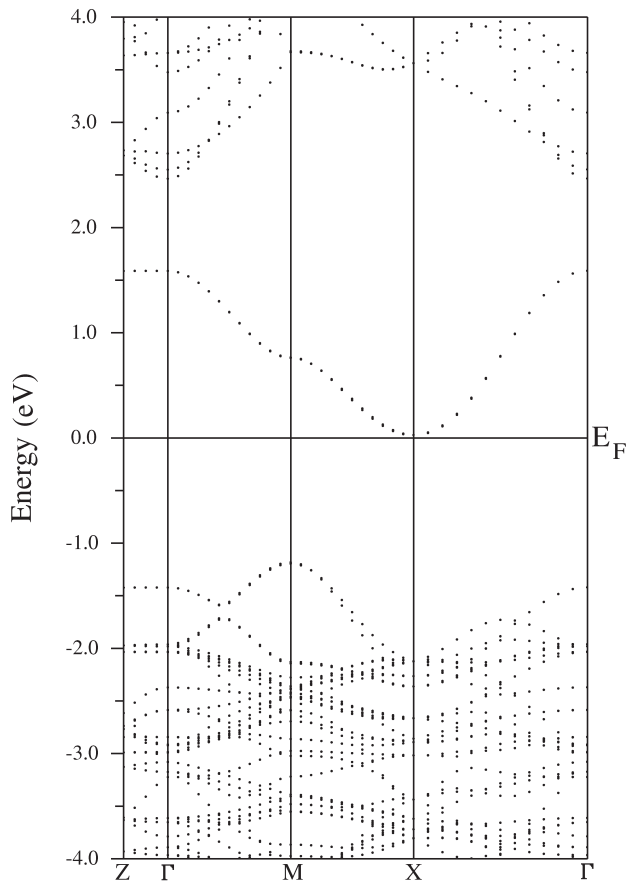}
\caption{The band structure of the insulating AFM
Sr$_{2}$CuO$_{2}$Cl$_{2}$. The Fermi energy is set to zero. Z is
(0, 0, $\pi$) and other high symmetry k-points are defined in Fig.
1.}\label{fig2}
\end{figure}

\begin{figure}
\centering
\includegraphics[width=0.47\textwidth]{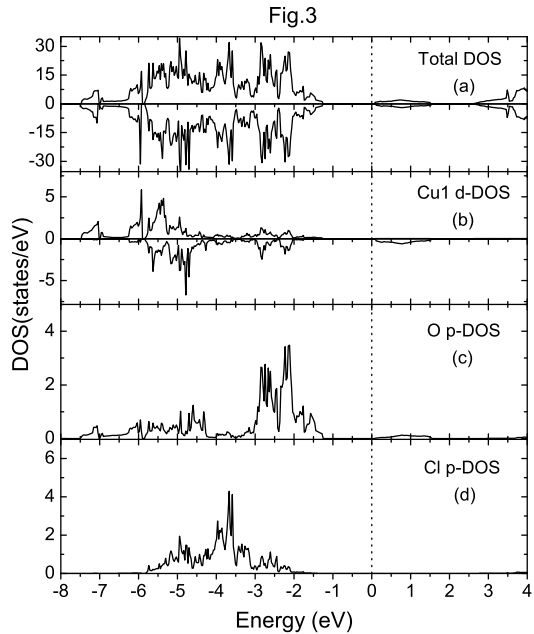}
\caption{Total and projected density of states (DOS) of AFM
Sr$_{2}$CuO$_{2}$Cl$_{2}$. Vertical dashed line refers to the
Fermi level. (a) Total DOS for spin up and down; (b) partial DOS
of Cu1 3d; (c) and (d) are DOS of O 2p and Cl 3p,
respectively.}\label{fig3}
\end{figure}

\begin{figure}
\centering
\includegraphics[width=0.47\textwidth]{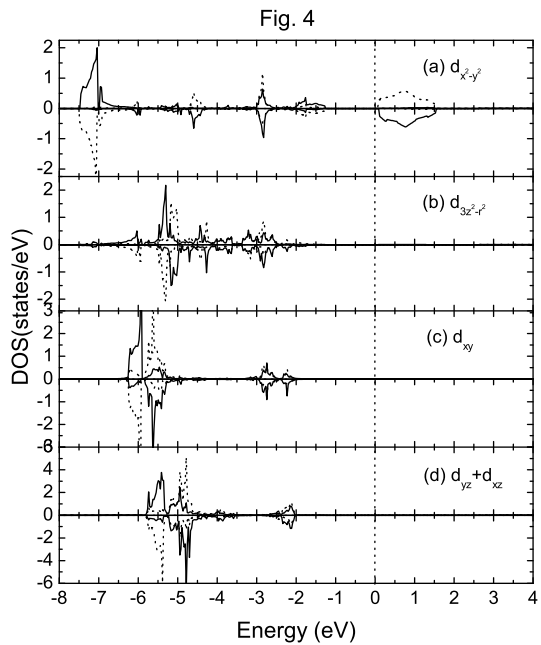}
\caption{Spin polarized projected DOS of Cu1 and Cu2 3d electrons.
Solid line represents that of Cu1 and dotted line, Cu2. Also the
vertical dashed line refers to Fermi level. (a) $d_{x^2-y^2}$, (b)
$d_{3z^2-r^2}$, (c) $d_{xy}$, (d)$d_{yz}+d_{xz}$.}\label{fig4}
\end{figure}

Although the hybridization of O 2p and Cu $3d_{x^2-y^2}$ between
-3 and -1.2 eV can be very clearly seen in Fig. 3(b) and 3(c), it
is still obvious that the topmost valence band is mostly composed
of the O 2p orbital, and a careful analysis shows that near the
top of the valence band the O 2p component is about 62.8{\%}.
Oppositely it is seen from Fig. 4(a) that the Cu $3d_{x^2-y^2}$
occupies more of the lowest conduction band with its component
being about 69.4{\%}. In the case of $U$=7.5 eV, the O 2p
component increases to 65.6{\%} at the top of the valence band and
Cu $3d_{x^2-y^2}$ occupies more to 70.6{\%} of the lowest
conduction band, which indicates that Sr$_{2}$CuO$_{2}$Cl$_{2}$ is
a charge transfer insulator. Thus the transition from O 2p to Cu
$3d_{x^2-y^2}$ orbitals will dominate the absorption edge. The Cl
3p orbital hybridizes with Cu 3d orbital in the range of from -5
to -3 eV, mostly with Cu $d_{yz}+d_{xz}$ and $d_{3z^2-r^2}$
orbitals (see Fig. 3(d), Fig. 4 (b) and (d)), all which are in the
plane perpendicular to the Cu-O plane. The hybridization between
Cl and Cu is less than that of Cu and O because O 2p orbitals
extends more than Cl 3p, and so hybridizes more with Cu 3d
orbitals.

\begin{figure}
\centering
\includegraphics[width=0.47\textwidth]{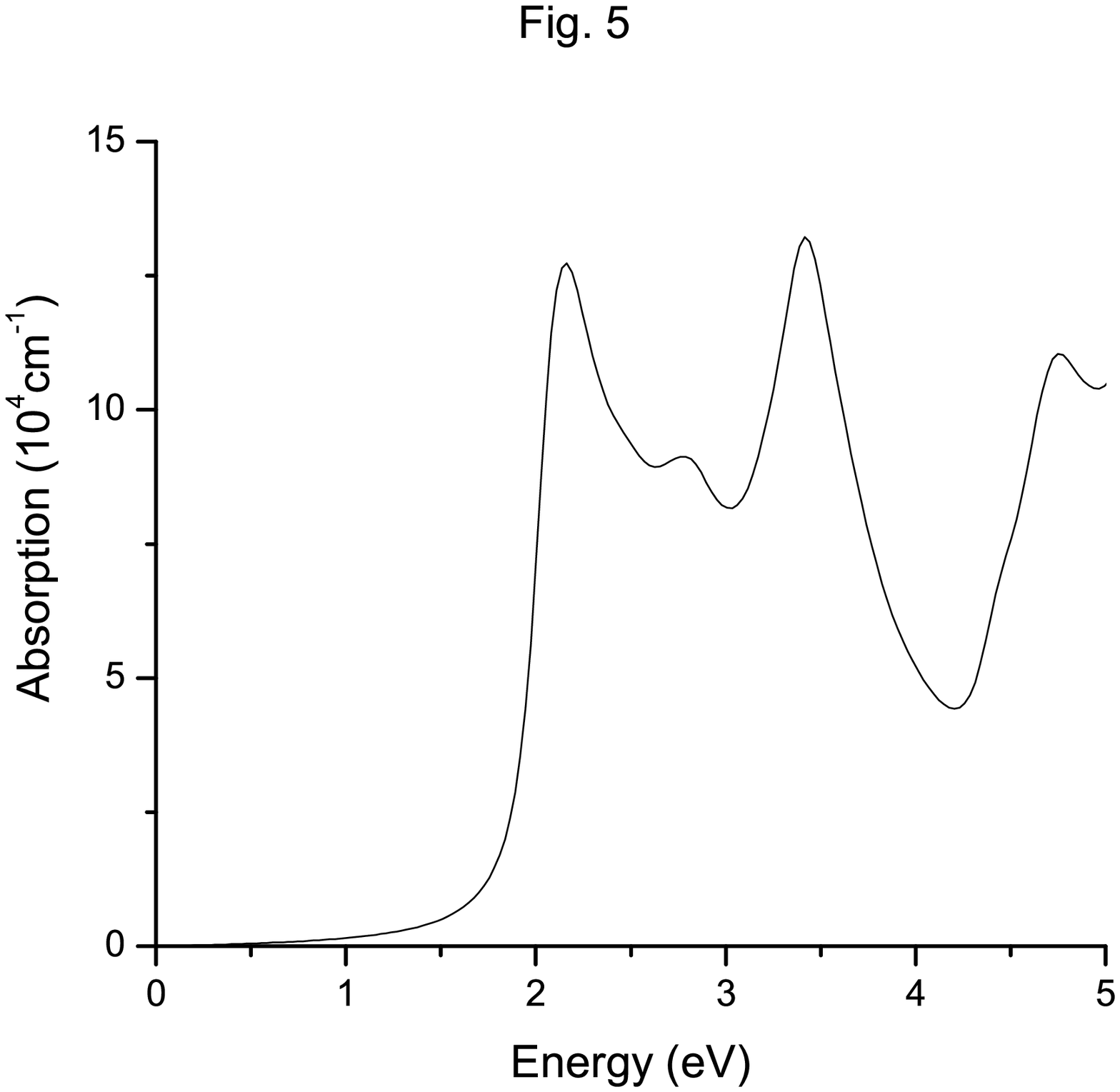}
\caption{The calculated absorption spectra of AFM
Sr$_{2}$CuO$_{2}$Cl$_{2}$.}\label{fig5}
\end{figure}

Study of optical property can give more informations on the
electronic structures. In Fig. 5, absorption spectrum calculated
with $U$=6.0 eV are presented. Although $U$=7.5 eV gives the
similar spectrum shapes, but all the peaks would shift to higher
energy by about 0.4 eV. So, we prefer to take $U$=6.0 eV as an
example in the following discussion. Comparing it with the
experimental data given by R. L\"{o}venich {\it et.al.}~\cite{13}
and Moskvin {\it et.al.}~\cite{15}, we find that they match very
well near the charge transfer gap of about 2eV, corresponding to
electrons transferred from O 2p to Cu 3d orbitals, which can be
clearly seen from the band structure in Fig. 2 along the path from
X($\pi $, 0) to M($\pi $/2, $\pi $/2). Around 2.5 eV, there is one
lower peak in both our calculation and Ref. [15], which was deemed
by Moskvin {\it et.al.}~\cite{15} as the other part of the
double-peak structure of the lowest excitation. The third strong
and broad peak appears at about 3.5 eV. According to the idea of
Pothuizen {\it et al.}~\cite{22}, a nonbonding band (NBB), mostly
O 2p$_\pi$ orbitals, lies at about 1.5 eV below the Zhang-Rice
band (ZRB), so the energy needed to transfer an electron from NBB
to Cu 3d is about 3.5 eV, which is also shown by Moskvin {\it et
al.} in their experimental observations. Again, the broad peak
around 4.6 eV appears both in our fist-principles result and their
measurements. Since the Cu $3d_{x^2-y^2}$ occupies the lowest part
of the conduction band from Fermi level to about 1.5 eV, and Cl
$3p$ dominates the most states between -4.5 and -2.5 eV, the quite
broad peak around 4.6 eV can be ascribed to the hopping of
electrons from Cl $3p$ orbitals to Cu $3d_{x^2-y^2}$ orbitals,
which can be clearly seen from Fig. 3 (b), (d) and Fig. 4 (a).
Usually, the CuO$_4$ model calculation on the
Sr$_{2}$CuO$_{2}$Cl$_{2 }$ has just neglected the Cl effect by
replacing the Cl atom by the O atom, while here we have shown the
contribution of Cl atoms to the optical properties of the
Sr$_{2}$CuO$_{2}$Cl$_{2 }$.

Finally, we want to emphasize that correct choice of $U$ value is critical for the LSDA+$U$ calculation. One criterion is whether the chosen $U$ value can give a reasonable physical result in consistent with the experimental observation. Combining our obtained results with that in Ref. [8] for $U$=5.0 eV, we can see that $U$=6.0 and 7.5 eV are both physically reasonable although the former has been selected for the main discussion on Sr$_{2}$CuO$_{2}$Cl$_{2}$'s optical properties. And it is clear that larger $U$ influences Cu $3d$ orbital more than O $2p$, and causes a wider gap, less band dispersion and a bigger magnetic moment on Cu. Also, larger $U$ shifts the absorption edge to higher energy.

\section{Conclusion} \label{Conclusion}

In summary, we have given a more detailed first-principle calculations on the band structure and linear optical properties of the AFM Sr$_{2}$CuO$_{2}$Cl$_{2 }$ by using the LSDA+$U$ approach in the FP-LAPW formalism. The obtained results indicate that it is a charge-transfer insulator with an indirect gap of about 1.2 eV, which is less than the experimental value. The $E$ {\it vs.} k relationship agrees very well to the ARPES observations. Also, we have found that our calculated linear optical susceptibility is well consistent with the experimental data not only around the charge-transfer gap, but also in the higher energy range.

\begin{acknowledgments}
The authors thank support to this work from a Grant for State Key
Program of China through Grant No. 1998061407. In addition, H. M.
Weng thanks a valuable discussion with Dr. Meichun Qian about the
FP-LAPW calculations. Our LSDA+$U$ calculation has been done on
the SGI origin 2000 Computer.
\end{acknowledgments}


\end{document}